# The Development of Educational Quality Administration: a Case of Technical College in Southern Thailand

Bangsuk Jantawan
Department of Tropical Agriculture and International Cooperation
National Pingtung University of Science and Technology
Pingtung, Taiwan
jantawan4@hotmail.co.th

Cheng-Fa Tsai
Department of Management Information Systems
National Pingtung University of Science and Technology
Pingtung, Taiwan
cftsai2000@yahoo.com.tw

*Abstract*— The purpose of this research were: to survey the needs of using the information system for educational quality administration; to develop Information System for Educational quality Administration (ISEs) in accordance with quality assessment standard; to study the qualification of ISEs; and to study satisfaction level of ISEs user. Subsequently, the tools of study have been employed that there were the collection of 47 questionnaires and 5 interviews to specialist by responsible officers for Information center of Technical colleges and Vocational colleges in Southern Thailand. The analysis of quantitative data has employed descriptive statistics using mean and standard deviation as the tool of measurement. Hence, the result was found that most users required software to search information rapidly (82.89%), software for collecting data (80.85%) and required Information system which could print document rapidly and ready for use (78.72%). The ISEs was created and developed by using Microsoft Access 2007 and Visual Basic. The ISEs was at good level with the average of 4.49 and SD at 0.5. Users' satisfaction of this software was at good level with the average of 4.36 and SD at 0.58.

I. INTRODUCTION

*A. Background*

According to the National Education Act (1999) and Vocational Education Act, 2008, the educational institution in Thailand had been changed in various aspect and also called for education reform. Education reform has been caused by social currents of globalization and knowledge. Not only Thailand, but also other countries around the world have focused on the teaching and learning process to students for holistic development. The concept of learning has made personnel of cultural, social, economic and technology for their continuous and timely global trends. The process of teaching and learning quality is important to make a difference. This process must be continuous and consistent with the concept of quality assurance to build up the confidence of students, parents, community and Thai society. Therefore, school administrators along with the teaching methods should confidently make standard ability of students and the impact on the development of Thailand [1], [2].

The section 47 in the National Education Act (1999) requires the development of quality assurance standards for all educational levels. It includes internal quality assurance and external quality assurance. In the section 48 requires the internal quality assurance as part of a process of the educational institutions continuously. In addition, the section 49 also requires that all educational institutions must obtain external assessment at least once in every five years since the last assessment. The assessment outcomes will be duly submitted to the concerned agencies and the public. Then, educational institutions must prepare to support implementation of the various sections [1],

Hence, researcher interests in development of information management system for administration of educational quality by collecting the data that related with the quality assurance standards. The Technical College and Vocational College in Southern Thailand have to improve a quality of information system and need to collect the data consistently in order to make the decision-maker efficiently. It can identify the weaknesses or problems effectively as well. The remedial measures are needed so as to facilitate subsequent planning and actions required to achieve the goals effectively [3].

*B. Research Objectives*

The overall objective of this study was to obtain information on the development of educational quality administration for the technical college in southern Thailand. There were four specific objectives of this study following: (1) to survey the needs of using information system for educational quality administration of colleges in southern Thailand; (2) to improve information system for educational quality administration of technical colleges in southern Thailand; (3) to study the quality of information system for educational quality administration of technical colleges in southern Thailand; and (4) to study satisfaction level of information system for educational quality administration of technical colleges in southern Thailand. Actions required achieving the goals effectively [3].

*C. Research Hypotheses*

This study was based on the hypotheses, which consists of the following three parts: (1) ISEs has appropriated in educational quality management of technical colleges in southern Thailand and good level; (2) ISEs has been developed and has a good quality; and (3) the users of ISEs have a satisfaction in the information system for educational quality management at a high level.

*D. Research Scope*

As the population, this study was created and develops ISEs of technical colleges in southern Thailand of 47 institutions. Purposive samplings composed of 47 responsible officers for information center of technical colleges and industrial and community colleges in southern Thailand.

As the specialist, there were 3 specialists of quality assessment information system for educational quality management. The details below were the specialist qualifications: (1) the graduate master's degree or higher than that of related fields; (2) the bachelor's degree or related work in the educational standards of technical colleges and vocational college; and (3) the working-related information and expertise in computer data base of not less than five years, including both sides of eight persons.

*E. Research Tools*

The tools of study enclosed of four parts: the first tool was the requirement questionnaire of ISEs software. The questionnaire divided was five episodes: (Episode 1) the general data of respondents were multiple choice questions; (Episode 2) the condition used information system in college was multiple choice questions; (Episode 3) the problem of used information system in college was multiple choice questions; (Episode 4) the need of used information system in college was multiple choice questions, and (Episode 5) the opinion and suggestion about needed in used information system program in college; the second tool was the ISEs software; the third tool was the quality evaluate of ISEs; and the fourth tool was the satisfaction questionnaire in used ISEs. It was divided were two parts: (Episode 1) the general data of respondents were multiple choice questions, and (Episode 2) the opinion about used information system was multiple choice questions.

*F. Definition Terminology*

There were 3 key terminologies of study. Firstly, the technical college in southern regional office of vocational commission: means the institutions technical college and vocational college kind within the technical college in southern regional office of vocational commission in the area south of 47 colleges. Secondly, the user information system: means the personal responsibility information center of technical college in southern regional office of vocational commission. Thirdly, the information system development: means the process of developing software with Microsoft Access and Visual Basic programming a standby Standalone to be able to store data, Process data and report data According to the information system.

II. METHODOLOGY

*A. Data Collection*

The process in this study consisted of the information following: (1) the data from questionnaires and the needs of using information system for educational quality administration of colleges in southern regional office of vocational education commission composed of 47 responsible officers; (2) the data from specialists of quality assurance standards, internal quality assurance, and overview information systems composed of eight specialists; and (3) the data from the technical college and vocational college in southern regional office of vocational commission composed of 47 institutions.

*B. Data Analysis*

*1) Create questionnaire needed in used information system for education quality administration;*

First step, there were 4 parts: (1) study the principle to create questionnaire for book document journal related research, (2) design and create survey of problem and needed to divide issues, (3) take the needed survey of used information system for education quality administration draft sent to chief advisor and chief advisor check it and Edit for suitable, (3) improve and edit needed survey used of information system for educational quality management of technical colleges in southern regional office of vocational education commission, and (4) take need survey in information system for education quality administration to use.

*2) Development for education quality administration development information system for education quality administration to create by procedure;*

Second step, there were 6 parts: (1) preliminary investigation, (2) systems analysis, (3) evaluate the consistency of the standard data items, (4) file and database design (file and database design, program structure design, and input and output design), (5) system development (programming, and documentation), and (6) system implementation.

*3) Creating evaluate for education quality administration have the procedure;*

Third step, there were 6 parts: (1) study the document ad related research to education quality administration, (2) research analysis form objective of research for approach in used evaluate for comprehensive objective of research, (3) create evaluate quality of system was evaluation scale 5 levels Likert scale for comments of the specialist of check system standard and tests information system before to use, (4) take standard evaluate sent to chiefs advisor to check and suggestion for suitable content, (5) improve and edit standard evaluate both side according suggestion, and (6) take quality evaluate of system to sent to the specialist 3 person quality evaluate.

*4) Satisfaction questionnaire of information system for education quality administration cerate to steps;*

Fourth step, there also were 6 parts: (1) study form book, document, journal and related research, (2) analysis research form aims of research for approach in create question to comprehensive of research, (3) design and create satisfaction questionnaire according analysis divided 2 parts such as part 1 general data of respondents, part 2 satisfactions of user in system, (4) take satisfactions evaluate draft sent to chief advisor and check it and edit for suitable content, (5) improve and edit evaluate satisfactions according suggestion, and (6) take satisfactions evaluate of user in information system to give personal user to evaluate.

### III. RESULTS

The demand survey of the information system for educational quality management of technical colleges in southern regional office of vocational education commission found that the majority of the information system users demand on quick access to information as a primary demand for 48.94 percent. The secondary demand is the quick search for information for 23.40 percent. Follow by the information arrangement in order for 14.89 percent, aggregate information for 6.38 percent, reduction of information redundancy for 4.26 percent, and reduction of information errors for 2.13 percent, respectively. The majority of the users demand on software that can quickly search the information related to their practices for 82.98 percent, the demand on information recording for 80.85 percent, the demand on software that support their reporting for 17.02 percent, and the demand on software that can reduce the information errors for 2.13, respectively. The users demand on the information system that is capable to print reports efficiently and quickly as a primary demand for 78.72 percent. The secondary demand is the report, produced by the information system, related to their practices for 14.89 percent. The format and details in each report must be easy to interpret for 6.38 percent.

The research results of the information system development for educational quality management of technical colleges in southern regional office of vocational education commission are divided into two parts. Part 1 is the summary of assessment analysis of the correlation between standard and data items with five experts, seven data standards, and 43 indicators. The summary of the validity of 43 questions in the questionnaire survey by five experts found that the correlation value is 1, indicated the correlation of all questions. Part 2 is the result of the system development, using data items that correlated with internal quality assessment standard for the information system structure design and development, including login interface, data input, and data display implemented by Visual Basic and Microsoft Access with the system size 11.7 megabytes.

The summary of the information system quality (Table I), educational quality management of technical colleges in southern regional office of vocational education commission by three experts found the good quality of the information system. According to the quality criteria (mean is 1.5 and standard deviation is 0.58). The quality in all dimensions are good according to the quality criteria, the quality of the data input is excellent (mean is 4.55 and standard deviation is 0.50), the quality of the results or reports is also excellent (mean is 4.5 and standard deviation is 0.58), the quality of the operational processes is good (mean is 4.30 and standard deviation is 0.50), and the quality of the content is also good (mean is 4.07 and standard deviation is 0.47).

The summary of the satisfaction assessment (Table II), information system using for educational quality management of technical colleges in southern regional office of vocational education commission are as follows: the satisfaction in the data input is high (mean is 4.38 and standard deviation is 0.58), the satisfaction in the contents is high (mean is 4.61 and standard deviation is 0.59), the satisfaction in the operational processes is also the high (mean is 4.63 and standard deviation is 0.61), and the highest satisfaction is in the results or reports (mean is 4.64 and standard deviation 0.53).

TABLE I. THE SUMMARY OF QUALITY ASSESSMENT FOR INFORMATION SYSTEM

| The Summary of Quality Assessment for Information System | $\bar{X}$ | S.D. | Quality Level |
|---|---|---|---|
| Quality of the data input | 4.55 | 0.50 | Excellent |
| Quality of the content | 4.07 | 0.47 | Good |
| Quality of the operational processes | 4.30 | 0.50 | Good |
| Quality of the results or reports | 4.50 | 0.58 | Excellent |
| **Average** | **4.36** | **0.58** | **Good** |

TABLE II. THE SUMMARY OF SATISFACTION ASSESSMENT FOR INFORMATION SYSTEM

| The Summary of Satisfaction Assessment for Information System | $\bar{X}$ | S.D. | Level of Satisfaction |
|---|---|---|---|
| Satisfaction of the data input | 4.38 | 0.58 | High |
| Satisfaction of the content | 4.61 | 0.59 | Very High |
| Satisfaction in the operational processes | 4.63 | 0.61 | Very High |
| Satisfaction is in the results or reports | 4.64 | 0.53 | Very High |
| **Average** | **4.52** | **0.52** | **Very High** |

### IV. CONCLUSIONS

The demands on the information system for educational quality management of technical colleges in southern regional office of vocational education commission found that the users, who have no experience with the information system for educational quality management, cause the disorder and redundancy information after they worked on it. The information system administrators with less experience want the training on the knowledge about information system. The activities on the information system still lack of supporting software, and this situation causes the problem in data recording that is difficult for the operations. The data search is slow and delaying works. For the results or information reporting, there is no information system that it can print out to quickly and approved for users' demand. The college that has complete, efficient, and up-to-date information system to serve the demands can improve the quality of that college efficiently. This development is based on the principles, evidences, and facts that can be proved by the scientific analyses and assessments, logics and causality because the information is required in planning and decision-making that leads to the development of concepts and alternative ways of operation [4].

The development of the information system for educational quality management of technical colleges in southern regional office of vocational education commission found that the correlation can be differentiated into seven dimensions as follows: learners and technical graduates, curriculum and study planning, learner development activities, professional services for publics, research & development, leadership and management, and the standard for the internal quality assurance. Generally, data items are correlated with the standard and the indicators in the internal educational quality assessment.

The information system users' satisfaction assessment for educational quality management of technical colleges in southern regional office of vocational education commission found that the highest satisfaction of the users is in the results or reports from the information system, high satisfaction in operational processes, contents, and data input.

The suggestions in this study consisted of the following items: Firstly, this study should have a management of information system, quality of education available online, not limited to only one computer at any computer to find information, and access data, anytime, anywhere; and Finally, database and information system of the college should be simple and easy to use, and actions required achieving the goals effectively and efficiently. Responsible officers should be maintaining the database and information system to effective and up date.

## V. SUGGESTIONS FOR FURTHER RESEARCH

Further work should be study the behavior of information systems personnel to use information system of the educational quality management of technical colleges in southern regional office of vocational education commission. The research with a similar format should be developing databases on information, report information, and software packages to more easily, such as MySQL database program or on-line system.

## ACKNOWLEDGMENT

B. Jantawan would like to express thanks Dr. Cheng-Fa Tsai, professor of the Management Information Systems Department, the Department of Tropical Agriculture and International Cooperation, National Pingtung University of Science and Technology in Taiwan for supporting the outstanding scholarship, and highly appreciates to the Technical College in Southern Thailand for giving the information.

## REFERENCES

[1] Office of the National Education Commission. (n.d.). National Education Act of B.E. 2542 (1999). Retrieved November 26, 2011, from http://www.onec.go.th/Act/5/english/act27.pdf

[2] Vocational Education Act, 2008, Retrieved November 30, 2011, from http://www.ratchakitcha.soc.go.th/DATA/PDF/2551/A/043/1.PDF

[3] Pinthong, C., 2010, Internal quality assurance standards for the College, and Community Colleges, Department typography Min Buri School.

[4] Boonreang, K., 1999, Statistical research1, Print No. 7, Bangkok: P.N, printing.